\documentclass[final,3p,times]{elsarticle}
\usepackage{amssymb}
\journal{Journal of Solid State Chemistry}

\begin{document}

\begin{frontmatter}


%%\title{Title\tnoteref{label1}}
%%\tnotetext[label1]{}

\title{Crystal Chemistry and Magnetic Properties of Manganese Zinc Alloy ``YMn$_2$Zn$_{20}$'' Comprising a Mn Pyrochlore Lattice}


%\author{Yoshihiko Okamoto\corref{cor1}, Takeshi Shimizu, Jun-ichi Yamaura, Yoko Kiuchi, and Zenji Hiroi}
\author{Yoshihiko~Okamoto\corref{cor1}}
\ead{yokamoto@issp.u-tokyo.ac.jp}
\author{Takeshi~Shimizu}
\author{Jun-ichi~Yamaura}
\author{Zenji~Hiroi}

\address{Institute for Solid State Physics, University of Tokyo, Kashiwa 277-8581, Japan}

\cortext[cor1]{Corresponding author}

\begin{abstract}
The chemical composition, crystal structure, and magnetic properties of a manganese zinc alloy with an ideal composition of YMn$_2$Zn$_{20}$, which comprises a pyrochlore lattice made of Mn atoms, are reported. The compound is stable only when In or Al is partially substituted for Zn. We have determined the actual chemical formula as YMn$_{2+\delta}$Zn$_{20-x-\delta}M_x$, with $M$ = In or Al, and have identified the characteristic preferences with which the incorporated $M$ and excess Mn$_{\delta}$ atoms occupy the three crystallographic sites for Zn atoms. The Mn atoms in the pyrochlore lattice possess small magnetic moments that interact with each other antiferromagnetically but exhibit no long-range order above 0.4 K, probably owing to the geometrical frustration of the pyrochlore lattice. As a result, the effective mass of the conduction electrons is considerably enhanced, as observed in the related pyrochlore-lattice compounds (Y,Sc)Mn$_2$ and LiV$_2$O$_4$. However, the presence of excess Mn atoms with large localized magnetic moments comparable to spin 5/2 tends to mask the inherent magnetism of the pyrochlore Mn atoms. It is suggested that ``YMn$_2$Zn$_{18}$In$_2$'' with neither excess Mn atoms nor site disorder would be an ideal compound for further study.
\end{abstract}


\end{frontmatter}

% \linenumbers

\section{Introduction}
\label{S1}
A family of ternary intermetallic phases with the general formula $AB_2C_{20}$ crystallizes in the cubic CeCr$_2$Al$_{20}$-type structure, where $A$ is a heavy element such as a rare earth, U, Ca, Zr, or Hf, $B$ is a transition metal, and $C$ is Zn or Al~\cite{1,2,3,4,5}. Many such compounds have been found to show a variety of electronic properties~\cite{5,B,C}. Recently, $AB_2C_{20}$ compounds with $A$ = Yb or Pr have attracted much attention because of unusual properties in their heavy fermion states. For example, YbCo$_2$Zn$_{20}$ shows an anomalously large Sommerfeld coefficient of 8 J K$^{-2}$ mol$^{-1}$~\cite{6}, while PrIr$_2$Zn$_{20}$ shows the coexistence of superconductivity and a quadrupole order associated with the non-Kramers doublet of Pr ions~\cite{7}. It is probably important for the emergence of these exotic phenomena that Yb and Pr ions with 4$f$ electrons are located in a high symmetry position, namely at the 8$a$ position in the space group $Fd\bar{3}m$ (Fig. 1).

The transition metal $B$ atoms occupy the 16$d$ positions and form a corner-sharing tetrahedral network called the pyrochlore lattice. Antiferromagnetic insulators with the pyrochlore lattice are known to show various unusual ground states such as a spin liquid, because a conventional N\'{e}el order can be suppressed by geometrical frustration~\cite{A}. On the other hand, there is a growing interest in metallic compounds that comprise pyrochlore lattices made of transition metal atoms in order to investigate possible effects of frustration on strongly correlated electrons. Among many metallic compounds having pyrochlore lattices, a cubic Laves phase (Y,Sc)Mn$_2$ and a spinel oxide LiV$_2$O$_4$ are fascinating, because they show heavy-fermion-like behavior, which is rare for $d$ electron compounds~\cite{8,9}. Since they have no $f$ electrons, the heavy-fermion states should not be caused by the Kondo effect between localized $f$ moments and conduction electrons as is ordinarily observed in Ce and Yb alloys. Alternatively, the formation mechanism may be ascribed to geometrical frustration of the pyrochlore lattice~\cite{10}, but is still under debate. Moreover, unlike in $f$-electron compounds, neither superconductivity nor other exotic ground states related to the heavy-fermion states have been discovered in $d$-electron compounds. A bottleneck for further investigation of the $d$-electron heavy-fermion state is the limited number of target compounds. We believe that there is a chance of finding new $d$-electron heavy-fermion compounds in the $AB_2C_{20}$ family. To date, however, most studies on these compounds have focused on the magnetic properties of the $A$ atoms, rather than on the $B$ atoms in the pyrochlore lattice. In order to study the magnetism of the $B$ atoms, a compound with a nonmagnetic $A$ atom such as Y is required. It is known from previous studies that Y$B_2C_{20}$ compounds show conventional Pauli paramagnetic behavior with no evidence of antiferromagnetic correlations; only YFe$_2$Zn$_{20}$ shows a magnetic response, although it lies in the vicinity of a ferromagnetic order~\cite{B,11}.

\begin{figure}[tb]
\begin{center}
\includegraphics[width=6cm]{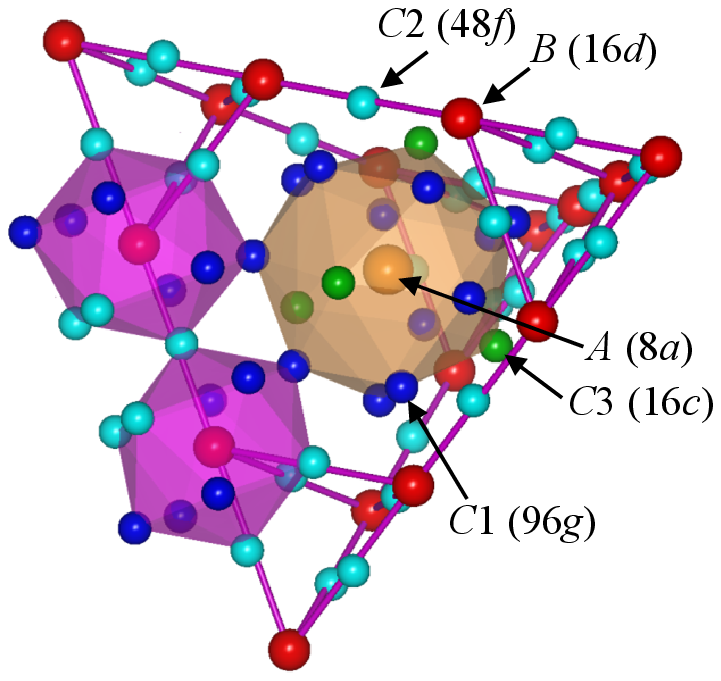}
\end{center}
\caption{(color online) Crystal structure of $AB_2C_{20}$ in the cubic CeCr$_2$Al$_{20}$ structure of space group $Fd\bar{3}m$. Coordination polyhedra surrounding an $A$ atom in the 8$a$ position and $B$ atoms in the 16$d$ positions are displayed. The $B$ atoms form a pyrochlore lattice as represented by thick bonds. Three kinds of $C$ atoms are labeled as $C$1, $C$2, and $C$3. The Wyckoff position for each atom is shown in parentheses. 
}
\label{F1}
\end{figure}

We have focused on $AB_2C_{20}$ with $A$ = Y and $B$ = Mn. Mn tends to show strong magnetism even in metallic compounds, because it favors a high-spin state in the half-closed-shell configuration with 3$d^5$ electrons. In the $AB_2C_{20}$ family, Mn compounds are rare, that is to say, chemically unstable compared with those of other 3$d$ compounds. One can obtain aluminum compounds $AB_2$Al$_{20}$ only for $B$ = Ti, V, and Cr, which are located to the left of Mn in the 3$d$ series of the periodic table, and zinc compounds $AB_2$Zn$_{20}$ only for $B$ = Fe, Co, and Ni, located to the right of Mn; the large Al atom prefers larger $B$ atoms, while the small Zn atom prefers smaller $B$ atoms. Thus, $AB_2C_{20}$ compounds with $B$ = Mn cannot be prepared with either $C$ = Al or $C$ = Zn. Recently, however, Benbow and Latturner succeeded in preparing ``YMn$_2$Zn$_{20}$'' by partially substituting In or Al for Zn~\cite{12}. 

In our previous study, we prepared In-substituted YMn$_2$Zn$_{20}$ and found strong Curie-Weiss magnetism originating from Mn 3$d$ electrons and a large heat capacity divided by temperature of 280 mJ K$^{-2}$ mol$^{-1}$ at 0.4 K, suggesting a significantly large mass-enhancement of conduction electrons~\cite{13}. However, there remain several open questions on the chemical and physical properties of the In-substituted YMn$_2$Zn$_{20}$. One concerns the fact that In-substituted YMn$_2$Zn$_{20}$ samples contain a considerable amount of excess Mn, i.e., more than 2 per formula unit~\cite{13}. The amount of excess Mn atoms and their influence on physical properties have yet to be determined. Furthermore, the properties of Al-substituted YMn$_2$Zn$_{20}$ have not yet been studied. In this paper, we report the chemical compositions, crystal structures, and physical properties of two series of single crystals of systematically controlled In and Al compositions. Characteristic evolutions of various properties with substitutions have been observed and are interpreted in terms of two distinct contributions from the Mn atoms in the pyrochlore lattice and from excess Mn atoms at the Zn sites. The former Mn atoms are weakly magnetic and may be responsible for the mass enhancement, while the latter Mn atoms show large local magnetic moments approximately corresponding to spin 5/2. 

\section{Experimental Procedure}
\label{S2}

\subsection{Synthesis}
\label{S2-1}
Two series of single crystals of In- and Al-substituted YMn$_2$Zn$_{20}$ were prepared by the melt-growth method. Y chips, Mn powder, Zn wires, and In or Al shots were mixed in a molar ratio of 1 : 2 : 20 $-$ $x_{\mathrm{n}}$ : $x_{\mathrm{n}}$. Seven In-substituted samples with nominal compositions of $x_{\mathrm{n}}$ = 1.5, 2, 3, 4, 5, 7, and 9 were prepared. The five samples with $x_{\mathrm{n}}$ = 3, 4, 5, 7, and 9 are the same as those reported in the previous study~\cite{13}. Six Al-substituted samples with $x_{\mathrm{n}}$ = 3, 5, 6, 7, 10, and 12 were prepared. Each mixture was placed in an alumina crucible and sealed in an evacuated quartz tube. The tube was heated to 900 $^{\circ}$C, kept at this temperature for 24 h, and then slowly cooled to 450 $^{\circ}$C over 225 h for the In-substituted samples and to 750 $^{\circ}$C over 150 h for the Al-substituted samples, followed by furnace cooling to room temperature. The crystals thus prepared were several mm in size and showed a metallic luster on their \{111\} habit faces, as shown in Fig. 2. Flux left on the surface of single crystals was mechanically removed. 

\begin{figure}[tb]
\begin{center}
\includegraphics[width=5cm]{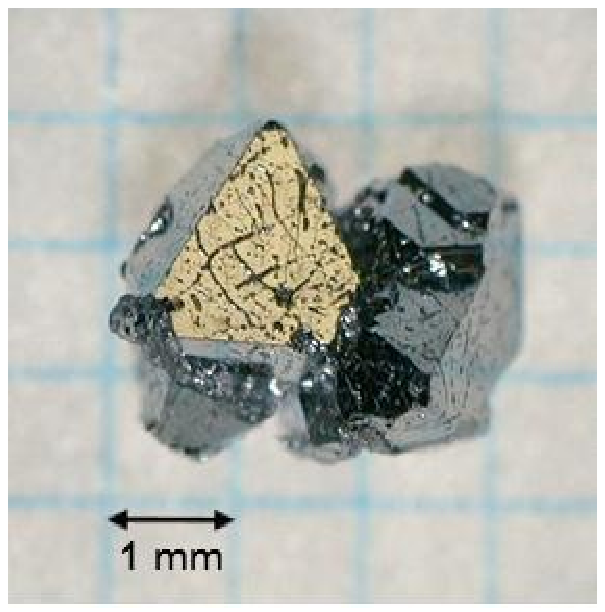}
\end{center}
\caption{(color online) A stereomicroscope image of single crystals of In-substituted YMn$_2$Zn$_{20}$ ($x$ = 3.46) on a grid of spacing 1 mm.
}
\label{F2}
\end{figure}

\subsection{Chemical and Structural Analyses}
\label{S2-2}
The chemical compositions of the In- and Al-substituted single crystals were determined by inductively-coupled plasma atomic-emission spectroscopy (ICP-AES) on a JY138KH apparatus (HORIBA). Samples for analysis were each prepared by dissolving a crystal weighing approximately 10 mg in dilute nitric acid. Quantitative analyses were performed by the matrix-matching method. Lattice constants at room temperature were determined by means of powder X-ray diffraction (XRD) using Cu-K$\alpha$ radiation on a RINT-2000 diffractometer (Rigaku). Data for structural analyses were collected at room temperature on a three-circle X-ray diffractometer equipped with a CCD detector (Bruker), using graphite-monochromated Mo-K$\alpha$ radiation. Structural parameters were refined by the full-matrix method using the Shelxl program~\cite{14} with XRD data up to a large diffraction angle of 2$\theta \sim$ 110$^{\circ}$. An absorption correction was applied by means of an analytical method~\cite{15}. The anisotropic displacement parameter and the occupancy of each site were accurately determined. 

\subsection{Physical Properties}
\label{S2-3}
Magnetic susceptibility measurements between 2 and 300 K on In- and Al-substituted YMn$_2$Zn$_{20}$ single crystals were made on a Magnetic Property Measurement System (Quantum Design). Heat capacity between 0.4 and 300 K for In-substituted samples was measured by the relaxation method using a crystal of approximately 1 mg on a Physical Property Measurement System (Quantum Design). Data between 0.4 and 2 K were obtained using a $^3$He refrigeration system.

\section{Results}
\label{S3}
\subsection{In-Substituted Samples}
\label{S3-1}
\subsubsection{Chemical Compositions and Lattice Constants}
\label{S3-1-1}

\begin{table}[tb]
\begin{center}
\caption{Chemical compositions and lattice constants of In-substituted samples. Nominal In content $x_{\mathrm{n}}$ and actual compositions determined by ICP-AES analyses for In ($x$), Zn ($y$), Mn (2 + $\delta$), and the sum ($x$ + $y$ + $\delta$) are listed for seven samples. 
}
\label{t1}
\footnotesize
\begin{tabular}{llllll}
& & & & & \\
\hline
$x_{\mathrm{n}}$ & $x$ (In) & $y$ (Zn) & 2 + $\delta$ (Mn) & $x$ + $y$ + $\delta$ & $a$ (\r{A}) \\
\hline
1.5 & 2.36(1) & 17.5(2) & 2.11(1) & 20.0(2) & 14.327(1) \\
2 & 2.39(1) & 17.47(4) & 2.23(1) & 20.09(6) & 14.3596(7) \\
3 & 2.96(3) & 16.64(5) & 2.44(2) & 20.0(1) & 14.4073(8) \\
4 & 3.22(1) & 16.2(1) & 2.64(2) & 20.1(1) & 14.417(1) \\
5 & 3.46(2) & 15.9(1) & 2.65(1) & 20.0(1) & 14.4659(7) \\
7 & 3.77(2) & 15.25(7) & 2.99(1) & 20.0(1) & 14.5024(5) \\
9 & 3.99(2) & 14.79(3) & 3.24(3) & 20.02(7) & 14.5273(3) \\
\hline
\end{tabular}
\end{center}
\end{table}

\begin{figure}[tb]
\begin{center}
\includegraphics[width=7cm]{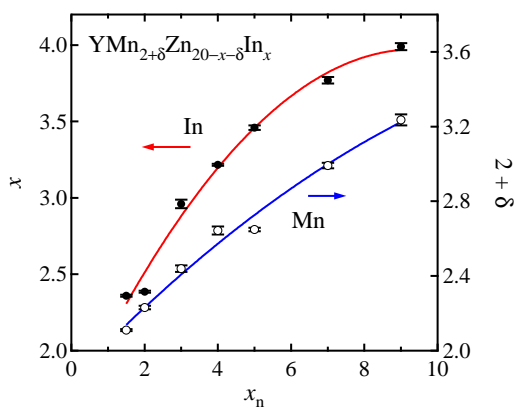}
\end{center}
\caption{(color online) Actual In content $x$ (left) and Mn content 2 + $\delta$ (right) determined by ICP-AES analyses plotted as a function of the nominal In content $x_{\mathrm{n}}$ for the In-substituted samples.
}
\label{F3}
\end{figure}

\begin{figure}[tb]
\begin{center}
\includegraphics[width=7cm]{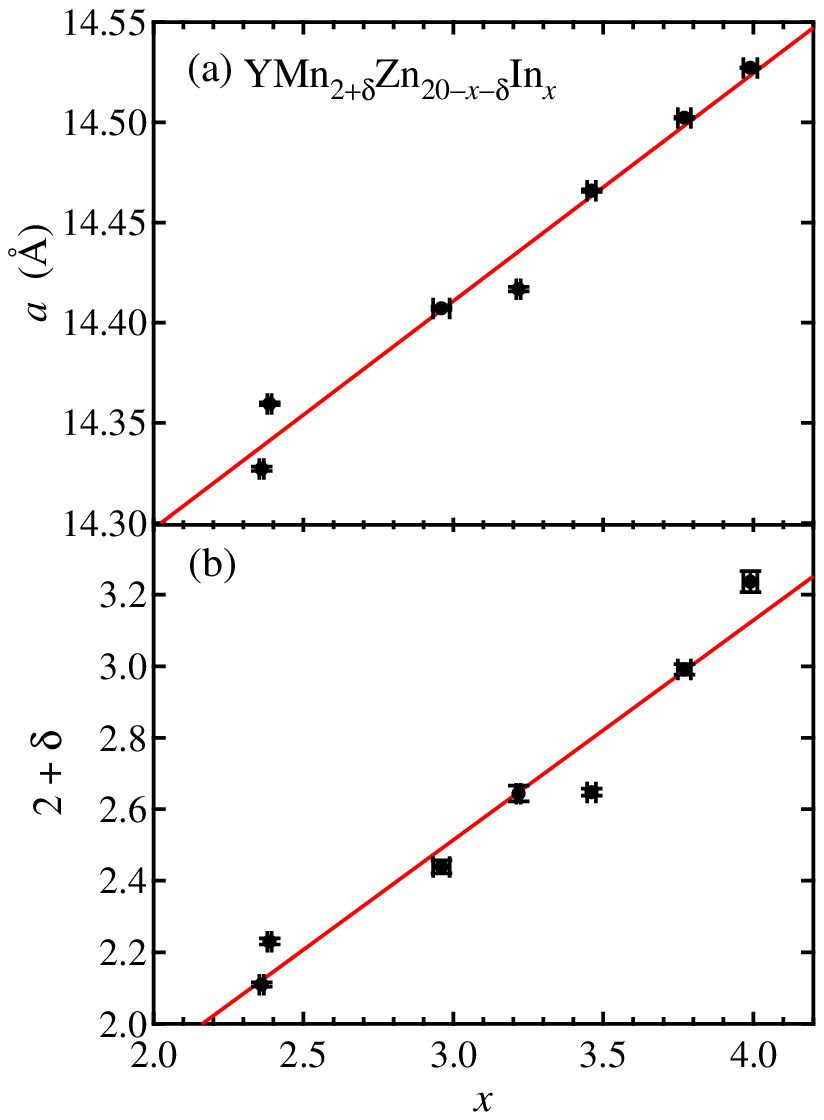}
\end{center}
\caption{(color online) Dependences on $x$ of lattice constant $a$ (a) and Mn composition 2 + $\delta$ (b) for In-substituted samples. The solid lines show linear fits, giving $a$ = 0.113(8)$x$ + 14.07(2) and 2 + $\delta$ = 0.61(6)$x$ + 0.6(19). 
}
\label{F4}
\end{figure}

The chemical compositions in terms of In ($x$), Zn ($y$), and Mn (2 + $\delta$) determined by ICP-AES analysis are listed in Table 1. All values are relative to that of Y. Since the 8$a$ position is fully occupied by Y atoms, as evidenced by structural refinements that will be mentioned later, these relative compositions should represent the actual content of each atom. Note that their sum is almost equal to 22 for every sample, indicative of an absence of vacancies at the other sites within the experimental resolution. Thus, the chemical formula of In-substituted YMn$_2$Zn$_{20}$ should be generally written as YMn$_{2+\delta}$Zn$_{20-x-\delta}$In$_x$.

The actual In content $x$ gradually increases with increasing nominal content $x_{\mathrm{n}}$, but is always different from $x_{\mathrm{n}}$, as compared in Table 1 and Fig. 3. For example, the $x_{\mathrm{n}}$ = 1.5 sample has a larger $x$ of 2.36, while the $x_{\mathrm{n}}$ = 9 sample has a smaller value of $x$ = 3.99. On the other hand, the actual Mn content is always greater than 2 for all of the samples; the excess content over 2 is denoted as $\delta$. As shown in Fig. 3, $\delta$ increases with increasing $x_{\mathrm{n}}$. For the smallest $x_{\mathrm{n}}$ = 1.5, $\delta$ is 0.11, which corresponds to 5\% of all of the Mn atoms, while for the largest $x_{\mathrm{n}}$ = 9, $\delta$ reaches 1.24. These excess Mn atoms must be located in the Zn sites together with In and Zn atoms, as ($x$ + $y$ + $\delta$) $\sim$ 20. The presence of excess Mn atoms in In-substituted YMn$_2$Zn$_{20}$ was not noted in a previous paper by Benbow and Latturner~\cite{12}.

Figure 4 shows lattice constant $a$ determined by means of powder XRD and Mn composition 2 + $\delta$ as a function of actual In content $x$. The value of $a$ is seen to increase linearly with increasing $x$, implying that the $x$ values obtained above are reliable. The positive slope of the line is consistent with the fact that the metallic radius of In is larger than that of Zn. Benbow and Latturner reported $a$ = 14.7285(4) \r{A} for a single crystal with $x$ = 5~\cite{12}, which is close to $a$ = 14.638 \r{A} obtained by extrapolating the line in Fig. 4(a) to $x$ = 5. On the other hand, $\delta$ also increases linearly with $x$, indicating that there is a correlation between the In and excess Mn contents; excess Mn atoms tend to be introduced at the Zn sites when In atoms occupy the Zn sites, as will be discussed later. 

\subsubsection{Crystal Structures}
\label{S3-1-2}

\begin{table*}[tb]
\begin{center}
\caption{Experimental and crystallographic data for four In-substituted samples with $x$ = 2.36, 2.96, 3.46, and 3.99.}
\label{t2}
\footnotesize
\begin{tabular}{lllll}
& & & & \\
\hline
& $x$ = 2.36 & 2.96 & 3.46 & 3.99 \\
\hline
$a$ (\r{A}) & 14.3344(1) & 14.4012(1) & 14.4658(1) & 14.5223(1) \\
2$\theta$ max ($^{\circ}$) & 106.4 & 106.6 & 106.7 & 106.8 \\
Total refs. & 7929 & 8202 & 8387 & 8439 \\
Independent refs. & 870 & 895 & 897 & 923 \\
Obs. refs. ($I >$ 3$\sigma$($I$)) & 863 & 810 & 771 & 841 \\
$R$(F) & 0.059 & 0.023 & 0.031 & 0.022 \\
$wR$(F$^2$) & 0.171 & 0.068 & 0.060 & 0.056 \\
\hline
\multicolumn{5}{l}{Space group $Fd\bar{3}m$ (No. 227), $Z$ = 8.}
\end{tabular}
\end{center}
\end{table*}

\begin{table*}[tb]
\begin{center}
\caption{Crystallographic parameters for four In-substituted YMn$_2$Zn$_{20}$ samples determined by means of single-crystal XRD. The occupancy at the Zn1 site was obtained by assuming only Zn atoms in this site. 
}
\label{t3}
\footnotesize
\begin{tabular}{lllllll}
& & & & & & \\
\hline
& Wyckoff position & $x$ & $y$ & $z$ & $U_{\mathrm{eq}}$/$U_{\mathrm{iso}}$ (\r{A}$^2$) & $g^*$ \\
\hline
$x$ = 2.36 \\
Y & 8$a$ & 1/8 & 1/8 & 1/8 & 0.0040(2) & 1 \\
Mn & 16$d$ & 1/2 & 1/2 & 1/2 & 0.0042(2) & 1 \\
In & 16$c$ & 0 & 0 & 0 & 0.0109(2) & 1 \\
Zn2 & 48$f$ & 0.49092(2) & 1/8 & 1/8 & 0.0080(2) & 1 \\
Zn1 & 96$g$ & 0.05703(1) & 0.05703(1) & 0.32842(2) & 0.0107(2) & 1.016(2) \\
\\
$x$ = 2.96 \\
Y & 8$a$ & 1/8 & 1/8 & 1/8 & 0.00356(8) & 1 \\
Mn & 16$d$ & 1/2 & 1/2 & 1/2 & 0.00316(7) & 1 \\
In & 16$c$ & 0 & 0 & 0 & 0.01202(7) & 1 \\
Zn2 & 48$f$ & 0.49097(2) & 1/8 & 1/8 & 0.00788(6) & 1 \\
Zn1 & 96$g$ & 0.05713(1) & 0.05713(1) & 0.32837(2) & 0.01076(6) & 1.052(2) \\
\\
$x$ = 3.46 \\
Y & 8$a$ & 1/8 & 1/8 & 1/8 & 0.0043(1) & 1 \\
Mn & 16$d$ & 1/2 & 1/2 & 1/2 & 0.0038(2) & 1 \\
In & 16$c$ & 0 & 0 & 0 & 0.01465(9) & 1 \\
Zn2 & 48$f$ & 0.49149(3) & 1/8 & 1/8 & 0.00873(7) & 1 \\
Zn1 & 96$g$ & 0.05722(2) & 0.05722(2) & 0.32839(2) & 0.01178(7) & 1.064(2) \\
\\
$x$ = 3.99 \\
Y & 8$a$ & 1/8 & 1/8 & 1/8 & 0.00601(8) & 1 \\
Mn & 16$d$ & 1/2 & 1/2 & 1/2 & 0.00527(7) & 1 \\
In & 16$c$ & 0 & 0 & 0 & 0.01820(8) & 1 \\
Zn2 & 48$f$ & 0.49192(2) & 1/8 & 1/8 & 0.01037(6) & 1 \\
Zn1 & 96$g$ & 0.05731(1) & 0.05731(1) & 0.32835(2) & 0.01354(6) & 1.089(2) \\
\hline
\multicolumn{7}{l}{$U_{\mathrm{eq}}$ = ($\Sigma_i\Sigma_j$$U_{ij}a^*_i$$a^*_j$$a_i$$a_j$)/2}
\end{tabular}
\end{center}
\end{table*}

\begin{figure}[!htb]
\begin{center}
\includegraphics[width=7cm]{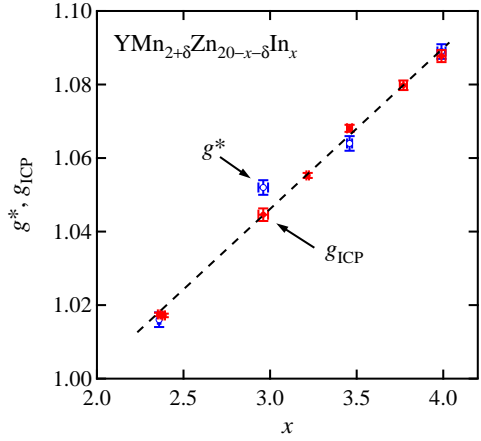}
\end{center}
\caption{(color online) Occupancies of the 96$g$ position for In-substituted samples. Plotted are $g^*$ determined by means of single-crystal XRD (open circle) and $g_{\mathrm{ICP}}$ from the data of ICP-AES analyses (filled circle). The broken line is a guide for the eyes.
}
\label{F5}
\end{figure}

\begin{table*}[tb]
\begin{center}
\caption{Selected bond lengths in \r{A} for In-substituted YMn$_2$Zn$_{20}$.}
\label{t4}
\footnotesize
\begin{tabular}{lllll}
& & & & \\
\hline
& $x$ = 2.36 & 2.96 & 3.46 & 3.99 \\
\hline
Y(8$a$)-In(16$c$) & 3.10349(1) & 3.11795(1) & 3.13194(1) & 3.14417(1) \\
Y(8$a$)-Zn1(96$g$) & 3.2251(2) & 3.2386(2) & 3.2526(3) & 3.2640(3) \\
Mn(16$d$)-Zn2(48$f$) & 2.53733(2) & 2.54912(2) & 2.5602(3) & 2.56988(2) \\
Mn(16$d$)-Zn1(96$g$) & 2.7177(3) & 2.7318(2) & 2.7446(3) & 2.7567(2) \\
In(16$c$)-Zn1(96$g$) & 3.0957(2) & 3.1089(3) & 3.1222(3) & 3.1333(2) \\
Zn2(48$f$)-Zn2(48$f$) & 2.7181(3) & 2.7297(3) & 2.7313(4) & 2.7331(3) \\
Zn2(48$f$)-Zn1(96$g$) & 2.7797(3) & 2.7947(2) & 2.8106(3) & 2.8248(2) \\
& 2.7064(2) & 2.7192(3) & 2.7367(3) & 2.7523(2) \\
Zn1(96$g$)-Zn1(96$g$) & 2.8400(3) & 2.8541(3) & 2.8665(4) & 2.8784(3) \\
 & 2.7558(2) & 2.7645(3) & 2.7743(4) & 2.7861(2) \\
 & 2.7458(2) & 2.7596(3) & 2.7733(4) & 2.7804(3) \\
\hline
\end{tabular}
\end{center}
\end{table*}

\begin{figure}[hbtp]
\begin{center}
\includegraphics[width=7cm]{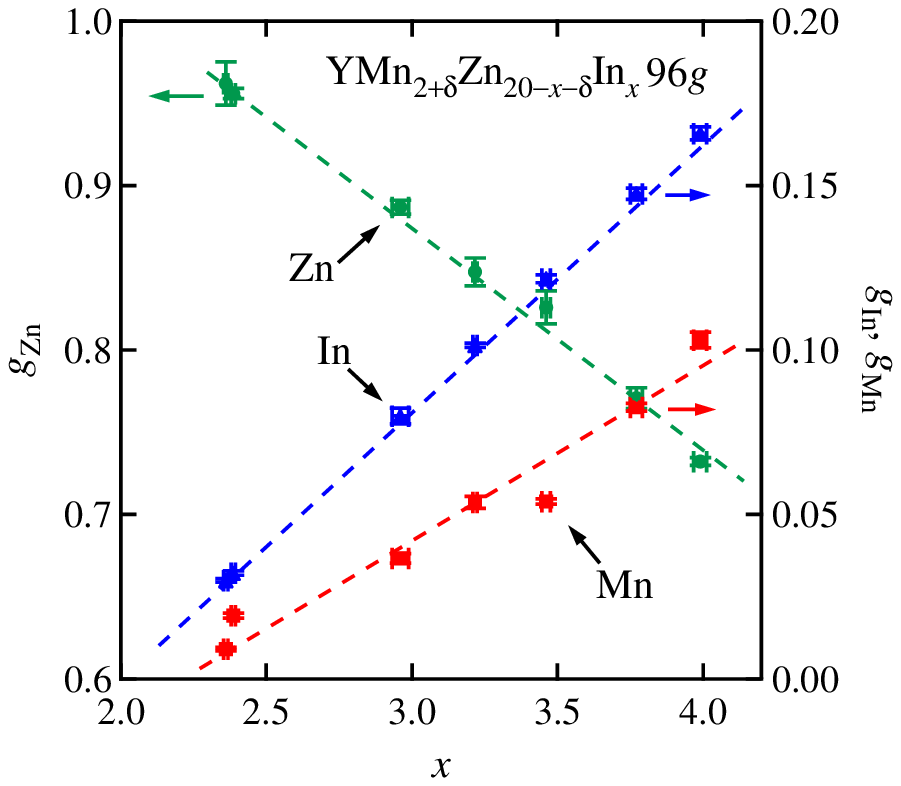}
\end{center}
\caption{(color online) Dependences on $x$ of the occupancies of the 96$g$ position by Zn (left axis), In (right axis), and Mn (right axis) for In-substituted samples, determined by ICP-AES analysis assuming 8$a$, 16$d$, 16$c$, and 48$f$ positions are occupied by Y, Mn, In, and Zn, respectively. 
}
\label{F6}
\end{figure}

Crystal structure refinements were carried out for four single crystals with $x$ = 2.36, 2.96, 3.46, and 3.99. The results and the obtained structural parameters are shown in Tables 2 and 3, and selected atomic distances are listed in Table 4. In the CeCr$_2$Al$_{20}$ structure of the space group $Fd$\=3$m$, the 8$a$ and 16$d$ positions are occupied by Ce and Cr atoms, respectively, while the 16$c$, 48$f$, and 96$g$ positions are occupied by Al atoms. According to the structural study on In-substituted YMn$_2$Zn$_{20}$ by Benbow and Latturner~\cite{12}, Y, Mn, and In are located at the 8$a$, 16$d$, and 16$c$ positions, respectively, while Zn occupies the 48$f$ and 96$g$ positions. We employed their model at the beginning of our refinements. However, we have noticed in the course of refinements that the occupancies of the 8$a$, 16$c$, 16$d$, and 48$f$ positions approach 1, while that of the 96$g$ position $g^*$ becomes significantly larger than 1. Thus, in the final refinement, only $g^*$ was refined with the others fixed at 1. As shown in Table 3, $g^*$ is slightly larger than 1 for all of the crystals. This means that other atoms heavier than Zn, i.e. Y or In, partially occupy the 96$g$ position. Since Y atoms are too large to fit this position, some In atoms must be present. 

The above results from structural refinements are consistent with those from ICP-AES analyses: In and excess Mn atoms larger than 2 are distributed over the Zn sites, i.e. $x$ + $y$ + $\delta$ = 20. The structural refinements show that two In atoms per chemical formula completely occupy the 16$c$ positions, and that the 48$f$ positions are fully occupied by Zn. Thus, ($x$ $-$ 2) In atoms should be located in the 96$g$ positions. Therefore, the chemical composition of the 96$g$ position must be represented as Zn$_{14-x-\delta}$In$_{x-2}$Mn$_{\delta}$. To confirm this, we compared the occupancies of the 96$g$ position determined in two ways: $g^*$ from structural refinements assuming that this position is occupied only by Zn atoms and $g_{\mathrm{ICP}}$ calculated assuming the chemical composition determined by ICP-AES analysis. Figure 5 shows the dependences of these occupancies on $x$. They are in surprisingly good agreement for all of the compositions, corroborating the above-mentioned site preferences. 

Figure 6 shows the dependences on $x$ of the occupancies of Zn, Mn, and In atoms in the 96$g$ position. For the smallest $x$ of 2.36, the 96$g$ position is 96\% occupied by Zn atoms, and only 1\% and 3\% occupied by Mn and In atoms, respectively. The occupancies of Mn and In atoms increase linearly with increasing $x$, and, correspondingly, that of Zn atoms becomes smaller. For the largest $x$ of 3.99, the occupancies of Mn and In atoms reach 10\% and 17\%, respectively. 

\begin{figure*}[hbtp]
\begin{center}
\includegraphics[height=4.5cm]{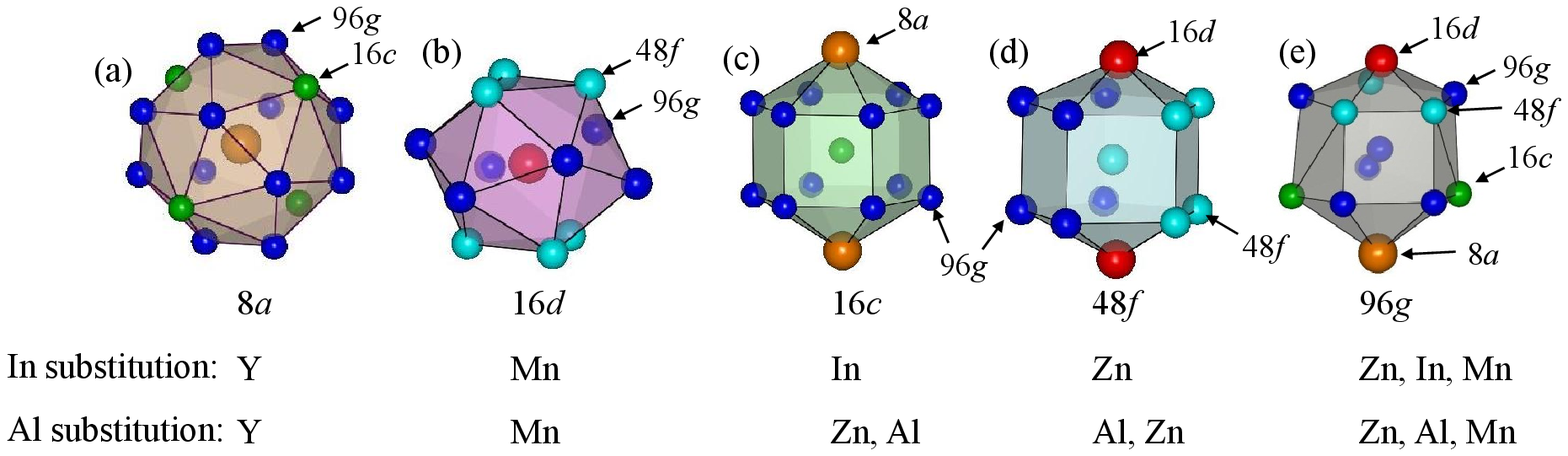}
\end{center}
\caption{(color online) Five coordinate polyhedra for In- and Al-substituted YMn$_2$Zn$_{20}$. Wyckoff positions and occupying atoms at the center of the polyhedra are listed below. }
\label{F7}
\end{figure*}

Figure 7 shows five kinds of coordination polyhedra that are present in the CeCr$_2$Al$_{20}$ structure. Site preferences observed for In-substituted YMn$_2$Zn$_{20}$ are shown at the bottom of the figure in addition to those for Al-substitution, which will be addressed later. As in other $AB_2C_{20}$ compounds, the 8$a$ position in a 16-coordinate polyhedron and the 16$d$ position in a 12-coordinate polyhedron are occupied by the most electropositive and largest $A$ atom (Y) and a transition metal $B$ atom (Mn), respectively. In contrast, the site preferences of the $C$ atoms (In, Mn, and Zn) over the 16$c$, 48$f$, and 96$g$ positions are characteristic in the present compound, and may be rationalized in terms of the atomic radii in metals~\cite{Muller}: 1.67, 1.37, and 1.34 \r{A} for In, Mn, and Zn, respectively. First, the 16$c$ position coordinated by 14 atoms is completely occupied by the largest In atom. This is reasonable because the bond lengths between the 16$c$ atom and ligands are considerably larger than those for the 48$f$ and 96$g$ positions, as shown in Table 4. In contrast, the 48$f$ position surrounded by a polyhedron comprising 12 atoms has the shortest bond lengths and is completely occupied by the smallest Zn atom. As a result, the remaining 96$g$ position in another medium-sized 12-coordinate polyhedron is randomly occupied by the rest of the Zn, In, and Mn atoms. 

\subsubsection{Magnetic Susceptibility}
\label{S3-1-3}
In this and the next sections, we discuss the physical properties of a series of In-substituted YMn$_2$Zn$_{20}$ single crystals by presenting magnetic susceptibility and heat capacity data. Magnetic susceptibilities measured in magnetic fields along the [100], [110], and [111] directions were almost the same, indicative of the absence of anisotropy as befits a cubic system. Figure 8 shows the temperature dependences of magnetic susceptibility for seven samples from $x$ = 2.36 to 3.99. All curves show Curie-Weiss behavior at high temperatures, indicating that a Mn atom has a localized magnetic moment or that a large spin fluctuation is present. At low temperatures approaching zero, they show saturating behavior. No such strong magnetic responses have hitherto been observed in other Y$B_2C_{20}$ compounds~\cite{B}; Mn atoms can retain ``magnetic moments'' even in a metallic compound, as expected. 

\begin{figure}[tb]
\begin{center}
\includegraphics[width=8cm]{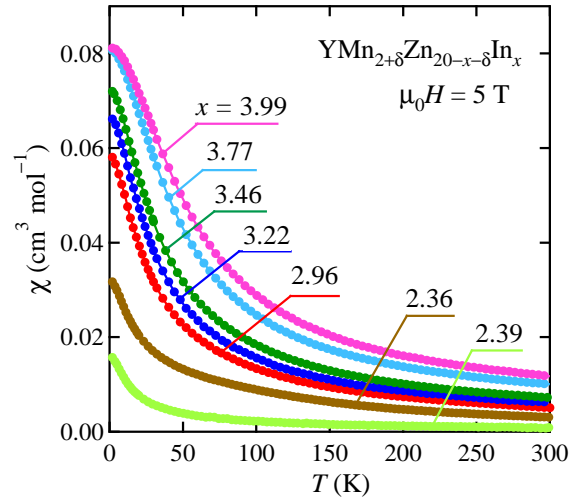}
\end{center}
\caption{(color online) Temperature dependence of magnetic susceptibility $\chi$ at a magnetic field of 5 T for In-substituted samples.}
\label{F8}
\end{figure}

\begin{figure}[!htb]
\begin{center}
\includegraphics[width=7cm]{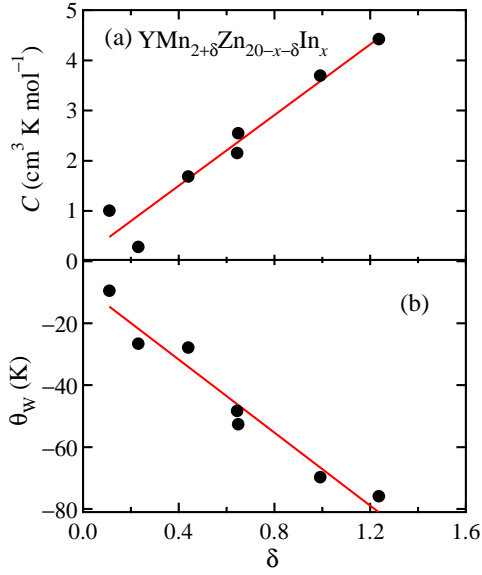}
\end{center}
\caption{(color online) Curie constant $C$ (a) and Weiss temperature $\theta_{\mathrm{W}}$ (b) obtained by Curie-Weiss fits to the magnetic susceptibility data shown in Fig. 8 plotted against excess Mn content $\delta$ for the In-substituted samples. The solid lines show linear fits to the data. 
}
\label{F9}
\end{figure}

\begin{figure}[tb]
\begin{center}
\includegraphics[width=8cm]{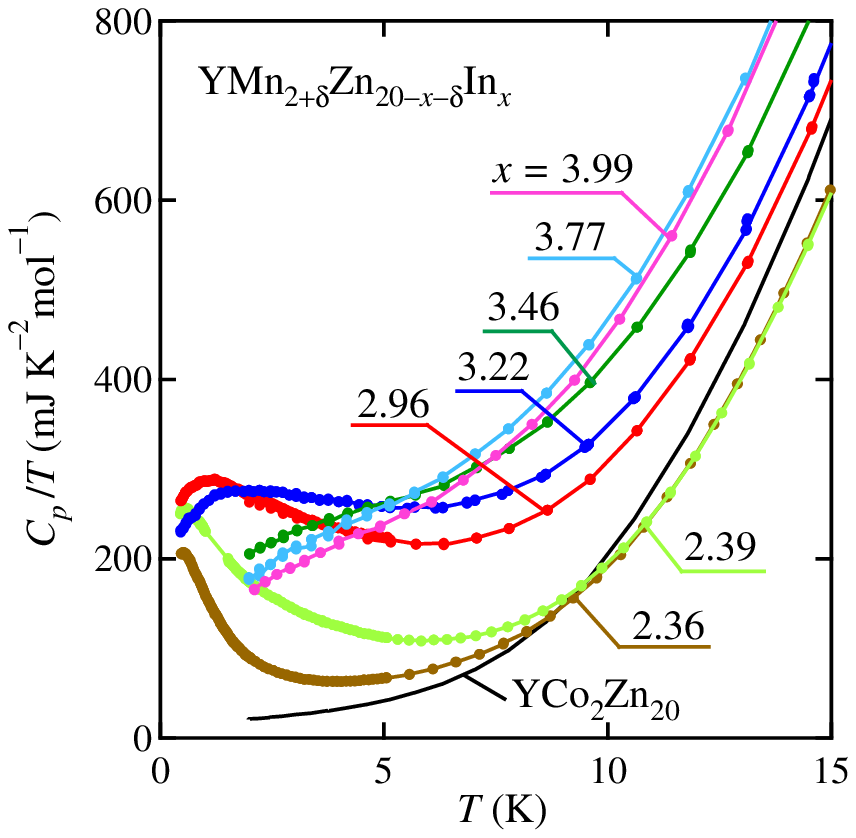}
\end{center}
\caption{(color online) Temperature dependence of heat capacity divided by temperature, $C_p$/$T$, for In-substituted samples. The data for a polycrystalline sample of YCo$_2$Zn$_{20}$ are also shown for comparison.
}
\label{F10}
\end{figure}

The magnitude of the magnetic susceptibility increases with increasing $x$, and, at the same time, the temperature dependence is enhanced. We fitted all the data in the temperature range 200-300 K to the Curie-Weiss form $\chi$ = $C$/($T$ $-$ $\theta_{\mathrm{W}}$) and thereby deduced Curie constants $C$ and Weiss temperatures $\theta_{\mathrm{W}}$, which are plotted as a function of excess Mn content $\delta$ in Fig. 9. Both $C$ and $\theta_{\mathrm{W}}$ are seen to vary linearly with $\delta$, indicating that $\chi$ is dominated by the excess Mn atoms in the 96$g$ position. A linear fit of the $C$ versus $\delta$ gives a slope of 3.5 cm$^3$ K mol$^{-1}$ and an intercept of 0.2 cm$^3$ K mol$^{-1}$ at $\delta$ = 0. This means that 1 mol of excess Mn atoms has a Curie constant of $C$ = 3.5 cm$^3$ K mol$^{-1}$, i.e., an effective magnetic moment of $\mu_{\mathrm{eff}}$ = 5.3 $\mu_{\mathrm{B}}$, where $\mu_{\mathrm{B}}$ is the Bohr magneton. This $\mu_{\mathrm{eff}}$ value is quite large, comparable to the $\mu_{\mathrm{eff}}$ = 5.92 $\mu_{\mathrm{B}}$ expected for spin-5/2 with $g$ = 2 in the 3$d^5$ configuration. The small intercept of 0.2 cm$^3$ K mol$^{-1}$ at $\delta$ = 0 must be ascribed to a contribution from the other Mn atoms in the 16$d$ position forming a pyrochlore lattice, which corresponds to an effective moment of approximately 1 $\mu_{\mathrm{B}}$, much smaller than that of an excess Mn atom. 

The Weiss temperatures in Fig. 9(b) are negative throughout, indicating that antiferromagnetic interactions are dominant. Since $\theta_{\mathrm{W}}$ decreases almost linearly with increasing $\delta$ and reaches a large value of $-$80 K at $\delta$ = 1.24 ($x$ = 3.99), strong antiferromagnetic interactions must exist between excess Mn spins; the likelihood of finding another Mn atom in the nearest-neighbor positions increases with increasing $\delta$. From the slope of a linear fit to the $\theta_{\mathrm{W}}$ data, that is $\theta_{\mathrm{W}}$/$\delta$ = $-$59 K, we estimated the magnitude of a nearest-neighbor interaction $J$ between excess Mn spins. The Weiss temperature is generally given by $\theta_{\mathrm{W}}$ = $-$($z$/3)$S$($S$ + 1)($J$/$k_{\mathrm{B}}$), where $z$ is the number of nearest-neighbor spins and the spin Hamiltonian is $H$ = $J$ $\Sigma_{i,j}\mathbf{S}_{i} \cdot \mathbf{S}_j$. In the present case, the number of 96$g$ positions surrounding a 96$g$ position, which is partly occupied by excess Mn atoms with a probability of $\delta$/12, is $z$ = 5 (see Fig. 7(e)). Provided that each excess Mn atom has spin-5/2 and that contributions from pyrochlore Mn spins at the 16$d$ position are negligible, $\theta_{\mathrm{W}}$ may be represented as $\theta_{\mathrm{W}}$ = $-$($z$/3)$S$($S$ + 1)($J$/$k_{\mathrm{B}}$)($\delta$/12) = $-$(175/144)($J$/$k_{\mathrm{B}}$)$\delta$. Thus, $\theta_{\mathrm{W}}$/$\delta$ = $-$59 K yields $J$/$k_{\mathrm{B}}$ = 49 K.

On the other hand, the intercept of the linear fit of $\theta_{\mathrm{W}}$ at $\delta$ = 0 in Fig. 9(b) is $-$11 K. This may correspond to a magnetic interaction between Mn spins in the pyrochlore lattice in the absence of excess Mn spins. Assuming $S$ = 1/2 for the pyrochlore Mn spins would give $J$/$k_{\mathrm{B}}$ $\sim$ 7 K ($z$ = 6), much smaller than the value deduced for excess Mn spins. This difference is readily understood by considering the interatomic distances between Mn atoms: that between pyrochlore Mn atoms is almost twice as large as those involving the excess Mn atoms; excess Mn atoms at the 96$g$ position are directly coordinated by themselves as shown in Fig. 7(e), while the pyrochlore Mn atoms at the 16$d$ position are intervened by Zn atoms at the 48$f$ position. 

As described above, an excess Mn atom randomly occupying the 96$g$ position has a large magnetic moment of $\sim$5 $\mu_{\mathrm{B}}$, while a pyrochlore Mn atom may have a small magnetic moment of $\sim$1 $\mu_{\mathrm{B}}$. Moreover, there always exist antiferromagnetic interactions between these spins. However, as shown in Fig. 8, no anomaly associated with long-range magnetic order is observed in any of the samples. This is probably because of geometrical frustration in the pyrochlore lattice as well as the random distribution of the excess Mn spins. Instead, magnetic susceptibility of the In-rich samples is strongly suppressed from a Curie-Weiss curve below a rather high temperature, for example, about 30 K for the $x$ = 3.99 sample. This must be due to antiferromagnetic short-range order between the excess Mn spins that may develop below $J$/$k_{\mathrm{B}}$ = 49 K. At lower temperature, the $x$ = 3.99 sample shows a thermal hysteresis in magnetic susceptibility below $T_{\mathrm{g}}$ = 10 K, indicating a spin-glass freezing transition~\cite{13}. $T_{\mathrm{g}}$ decreases with decreasing $\delta$. For $x$ $\le$ 3.22, no thermal hysteresis appears above 2 K. Therefore, this spin freezing is caused by randomly distributed excess Mn spins. The magnetism of the pyrochlore Mn spins is hard to observe because of the limited temperature range in the present experiments. 

\subsubsection{Heat Capacity}
\label{S3-1-4}

Figure 10 shows the temperature dependence of heat capacity divided by temperature $C_p$/$T$. That of a Pauli paramagnetic metal YCo$_2$Zn$_{20}$ is also shown for comparison. The $x$ = 3.99 sample, which has the largest $x$ and $\delta$, is seen to exhibit much larger $C_p$/$T$ than YCo$_2$Zn$_{20}$. This large additional contribution decreases and shifts to lower temperatures as $x$ and $\delta$ decrease. Thus, it is ascribed to magnetic heat capacity from excess Mn spins. The shift to lower temperature with decreasing $x$ is due to decreasing effective magnetic interactions, as shown in the dependence on $\delta$ of the Weiss temperature in Fig. 9(b); the magnetic entropy is released at lower temperatures as magnetic interactions are reduced. 
A broad peak in $C_p$/$T$ at 1.2 and 2 K for $x$ = 2.96 and 3.22, respectively, must be related to the spin-glass transition caused by excess Mn spins, as discussed in our previous study~\cite{13}. $x$ = 2.36 data do not show
such a broad peak down to 0.4 K, suggestive of absence of the spin-glass freezing in the $x$ = 2.36 sample.

The $C_p$/$T$ of the $x$ = 2.36 sample with the smallest $x$ and $\delta$ is close to that of YCo$_2$Zn$_{20}$ above 5 K, indicating no spin entropy of Mn 3$d$ electrons but lattice contributions. Below 5 K, $C_p$/$T$ increases rapidly upon cooling, saturating at a large value of 207 mJ K$^{-2}$ mol$^{-1}$ at 0.4 K. Since $\delta$ = 0.11 in this sample, corresponding to about 5\% of all Mn atoms, the contribution from the excess Mn must be very small, with that of the pyrochlore Mn being dominant. Thus, the observed large enhancement in $C_p$/$T$ below 5 K is seemingly mostly due to the spin entropy of pyrochlore Mn atoms. If so, this means that conduction electrons become heavy with a large Sommerfeld coefficient of 207 mJ K$^{-2}$ mol$^{-1}$, as observed in (Y,Sc)Mn$_2$ and LiV$_2$O$_4$~\cite{8,9}. Compared with $\gamma_{\mathrm{band}}$ = 31 mJ K$^{-2}$ mol$^{-1}$ for $x$ = 2 obtained by Harima in his band structure calculations~\cite{Harima}, there is a more than sixfold mass enhancement in the $x$ = 2.36 sample. However, this is only a rough estimation because of ambiguity in estimating the contribution from still present excess Mn spins. Further experiments on a $\delta$ = 0 sample or another experiment using a site-selective probe such as NMR would be required to clarify this important issue. 

\subsection{Al-Substituted Samples}
\label{S3-2}
\subsubsection{Chemical Compositions and Lattice Constants}
\label{S3-2-1}

\begin{table}[tb]
\begin{center}
\caption{Chemical compositions and lattice constants of Al-substituted samples. Nominal Al content $x_{\mathrm{n}}$ and actual compositions determined by ICP-AES analyses for Al ($x$), Zn ($y$), Mn (2 + $\delta$), and the sum ($x$ + $y$ + $\delta$) are listed for six samples. 
}
\label{t5}
\footnotesize
\begin{tabular}{llllll}
& & & & & \\
\hline
$x_{\mathrm{n}}$ & $x$ (Al) & $y$ (Zn) & 2 + $\delta$ (Mn) & $x$ + $y$ + $\delta$ & $a$ (\r{A}) \\
\hline
3 & 5.47(5) & 13.96(1) & 2.59(1) & 20.02(7) & 14.1564(8) \\
5 & 6.04(5) & 13.57(4) & 2.40(1) & 20.0(1) & 14.1550(2) \\
6 & 6.37(1) & 13.50(2) & 2.13(2) & 20.00(5) & 14.1669(5) \\
7 & 7.71(7) & 12.24(8) & 2.06(1) & 20.0(2) & 14.1769(5) \\
10 & 10.02(9) & 9.61(6) & 2.37(1) & 20.0(2) & 14.2210(4) \\
12 & 11.45(2) & 8.34(2) & 2.23(1) & 20.02(5) & 14.2413(5) \\
\hline
\end{tabular}
\end{center}
\end{table}

\begin{figure}[tb]
\begin{center}
\includegraphics[width=7cm]{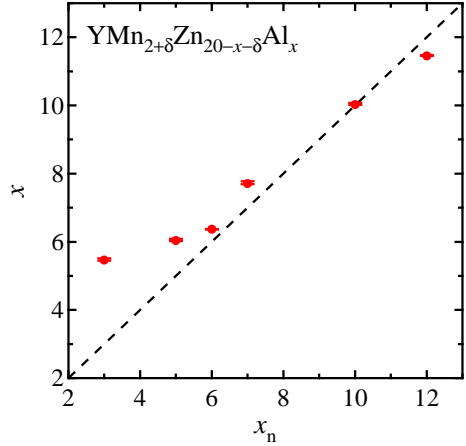}
\end{center}
\caption{(color online) Actual Al content $x$ determined by ICP-AES analyses as a function of the nominal Al content $x_{\mathrm{n}}$ for Al-substituted samples. The $x$ = $x_{\mathrm{n}}$ line is shown by a broken line.
}
\label{F11}
\end{figure}

\begin{figure}[!htb]
\begin{center}
\includegraphics[width=7cm]{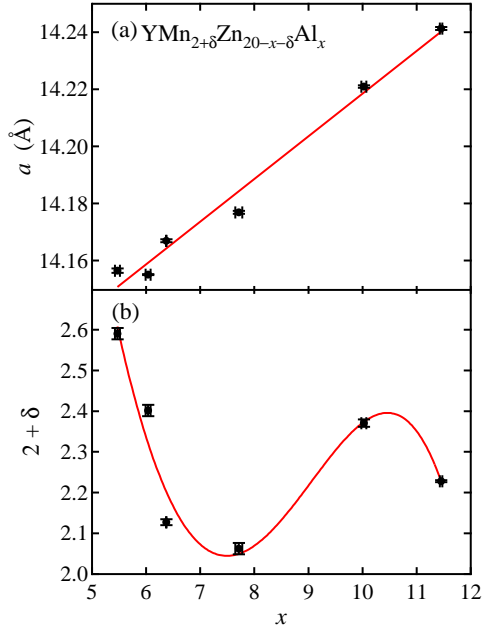}
\end{center}
\caption{(color online) Dependences on $x$ of lattice constant $a$ (a) and Mn composition 2 + $\delta$ (b) for Al-substituted samples. The solid line in (a) shows a linear fit, and the line in (b) is a guide to the eyes.
}
\label{F12}
\end{figure}

From ICP-AES analyses of six single-crystalline samples with nominal Al contents of $x_{\mathrm{n}}$ = 3, 5, 6, 7, 10, and 12, the chemical formula of Al-substituted YMn$_2$Zn$_{20}$ was established as YMn$_{2+\delta}$Zn$_{20-x-\delta}$Al$_x$. Thus, excess Mn atoms are present and occupy the Zn sites together with Al atoms, as in the In-substituted samples. Actual Al content $x$, Zn content $y$, and excess Mn content $\delta$ are listed in Table 5. As shown in Fig. 11, $x$ increases monotonically with increasing $x_{\mathrm{n}}$. $x$ is significantly larger than $x_{\mathrm{n}}$ for $x_{\mathrm{n}}$ = 3 and 5, while in the $x_{\mathrm{n}}$ = 6, 7, 10, and 12 samples, $x$ is almost same as $x_{\mathrm{n}}$.

Lattice constant $a$ determined by powder XRD measurements is plotted against $x$ in Fig. 12(a). The value of $a$ increases linearly with increasing $x$, consistent with the fact that the metallic radius of Al is larger than that of Zn. Benbow and Latturner reported $a$ = 14.1533(3) \r{A} for a single crystal with $x$ = 3.9~\cite{12}, which is close to the value of $a$ = 14.127 \r{A} obtained by extrapolating the line in Fig. 12(a) to $x$ = 3.9. 

\begin{table}[tb]
\begin{center}
\caption{Experimental and crystallographic data for three Al-substituted samples of $x$ = 6.04, 7.71, and 10.02.
}
\label{t6}
\footnotesize
\begin{tabular}{llll}
& & & \\
\hline
& $x$ = 6.04 & 7.71 & 10.02 \\
\hline
$a$ (\r{A}) & 14.1531(1) & 14.1815(1) & 14.2245(1) \\
2$\theta$ max ($^{\circ}$) & 106.5 & 106.2 & 106.7 \\
Total refs. & 7854 & 7607 & 7978 \\
Independent refs. & 863 & 864 & 875 \\
Obs. refs. ($I$ $>$ 3$\sigma$($I$)) & 803 & 821 & 800 \\
$R$(F) & 0.024 & 0.046 & 0.022 \\
$wR$(F$^2$) & 0.067 & 0.099 & 0.054 \\
\hline
\multicolumn{4}{l}{Space group $Fd\bar{3}m$ (No. 227), $Z$ = 8.}
\end{tabular}
\end{center}
\end{table}

\begin{table*}[tb]
\begin{center}
\caption{Crystallographic parameters for three Al-substituted YMn$_2$Zn$_{20}$ samples determined by means of single-crystal XRD. The occupancies at the Zn1, Zn2 and Zn3 sites were obtained by assuming only Zn atoms in these sites.
}
\label{t7}
\footnotesize
\begin{tabular}{lllllll}
& & & & & & \\
\hline
& Wyckoff position & $x$ & $y$ & $z$ & $U_{\mathrm{eq}}$/$U_{\mathrm{iso}}$ (\r{A}$^2$) & $g^*$ \\
\hline
$x$ = 6.04 & & & & & & \\
Y & 8$a$ & 1/8 & 1/8 & 1/8 & 0.00455(8) & 1 \\
Mn & 16$d$ & 1/2 & 1/2 & 1/2 & 0.00528(8) & 1 \\
Zn3 & 16$c$ & 0 & 0 & 0 & 0.0171(2) & 0.890(2) \\
Zn2 & 48$f$ & 0.48553(4) & 1/8 & 1/8 & 0.0095(1) & 0.514(2) \\
Zn1 & 96$g$ & 0.06000(1) & 0.06000(1) & 0.32321(2) & 0.01107(6) & 0.956(2) \\
& & & & & & \\
$x$ = 7.71 & & & & & & \\
Y & 8$a$ & 1/8 & 1/8 & 1/8 & 0.0040(2) & 1 \\
Mn & 16$d$ & 1/2 & 1/2 & 1/2 & 0.0049(1) & 1 \\
Zn3 & 16$c$ & 0 & 0 & 0 & 0.0176(4) & 0.776(8) \\
Zn2 & 48$f$ & 0.48508(9) & 1/8 & 1/8 & 0.0097(2) & 0.480(4) \\
Zn1 & 96$g$ & 0.06009(4) & 0.06009(4) & 0.32318(3) & 0.0109(1) & 0.938(6) \\
& & & & & & \\
$x$ = 10.02 & & & & & & \\
Y & 8$a$ & 1/8 & 1/8 & 1/8 & 0.00463(6) & 1 \\
Mn & 16$d$ & 1/2 & 1/2 & 1/2 & 0.00606(7) & 1 \\
Zn3 & 16$c$ & 0 & 0 & 0 & 0.0184(2) & 0.611(2) \\
Zn2 & 48$f$ & 0.48518(4) & 1/8 & 1/8 & 0.0102(1) & 0.451(2) \\
Zn1 & 96$g$ & 0.06001(1) & 0.06001(1) & 0.32336(1) & 0.01095(5) & 0.820(2) \\
\hline
\multicolumn{7}{l}{$U_{\mathrm{eq}}$ = ($\Sigma_i\Sigma_j$$U_{ij}a^*_i$$a^*_j$$a_i$$a_j$)/2}
\end{tabular}
\end{center}
\end{table*}

Compared with the simple dependence on $x$ of the lattice constant, the dependence on $x$ of the Mn composition 2 + $\delta$ is complicated, as shown in Fig. 12(b). With increasing $x$, $\delta$ first decreases from the largest value of 0.59 at $x$ = 5.47 to the smallest value of $\delta$ = 0.06 at $x$ = 7.71, but then increases once more with further increasing $x$. This is in contrast to the almost linear dependence on $x$ of $\delta$ for the In-substituted samples (Fig. 4). This difference appears to be ascribed to different site preferences of the incorporated atoms between the two series of samples, as will be described in the next section. The range of $\delta$ in the Al-substituted samples is between 0.06 and 0.59, which is narrower than that in the In-substituted samples. Moreover, since $\delta$ = 0.06 is almost half of the lowest $\delta$ value in the In-substituted samples, the influence of excess Mn atoms on physical properties is expected to be smaller in the Al-substituted samples. 

\subsubsection{Crystal Structures}
\label{S3-2-2}

Structural refinements were carried out for the $x$ = 6.04, 7.71, and 10.02 samples; a summary of the crystal data is given in Table 6, and the obtained structural parameters are listed in Table 7. As in the case of In substitution, the 8$a$ and 16$d$ positions are completely occupied by Y and Mn atoms, respectively. A difference from the In substitution is found in different occupations of the Zn sites by the incorporated Al: In atoms prefer the 16$c$ position, while Al atoms are distributed over all of the Zn sites. This is evidenced by the fact that the occupancies of the 16$c$, 48$f$, and 96$g$ positions are significantly smaller than unity when analyzed assuming that only Zn atoms are present. For example, in the $x$ = 7.71 sample, which contains the least excess Mn of $\delta$ = 0.06, the occupancies of the 16$c$, 48$f$, and 96$g$ positions are 0.776(8), 0.480(4), and 0.938(6) (Table 6), implying that lighter Al atoms replace Zn atoms particularly in the former two sites. A small amount of excess Mn atoms must also be present at these sites, although it was difficult to decide the occupation of the Mn atoms. We assume that they occupy only the 96$g$ positions as in the case of the In-substituted samples. 

\begin{table}[tb]
\begin{center}
\caption{Selected bond lengths in \r{A} for Al-substituted YMn$_2$Zn$_{20}$.}
\label{t8}
\footnotesize
\begin{tabular}{llll}
& & & \\
\hline
& $x$ = 6.04 & 7.71 & 10.02 \\
\hline
Y(8$a$)-Zn3(16$c$) & 3.06424(1) & 3.07038(1) & 3.07969(1) \\
Y(8$a$)-Zn1(96$g$) & 3.0923(3) & 3.0973(5) & 3.1097(1) \\
Mn(16$d$)-Zn2(48$f$) & 2.5103(4) & 2.5159(9) & 2.52338(5) \\
Mn(16$d$)-Zn1(96$g$) & 2.7754(3) & 2.7821(6) & 2.7876(1) \\
Zn3(16$c$)-Zn1(96$g$) & 3.0043(2) & 3.0094(6) & 3.0201(1) \\
Zn2(48$f$)-Zn2(48$f$) & 2.7916(6) & 2.806(1) & 2.8127(6) \\
Zn2(48$f$)-Zn1(96$g$) & 2.7943(3) & 2.8000(6) & 2.8069(2) \\
& 2.6401(3) & 2.639(1) & 2.6472(5) \\
Zn1(96$g$)-Zn1(96$g$) & 2.8894(3) & 2.8958(7) & 2.9016(2) \\
& 2.6663(3) & 2.6728(7) & 2.6829(2) \\
& 2.6020(3) & 2.6036(7) & 2.6147(2) \\
\hline
\end{tabular}
\end{center}
\end{table}

\begin{figure}[tb]
\begin{center}
\includegraphics[width=7cm]{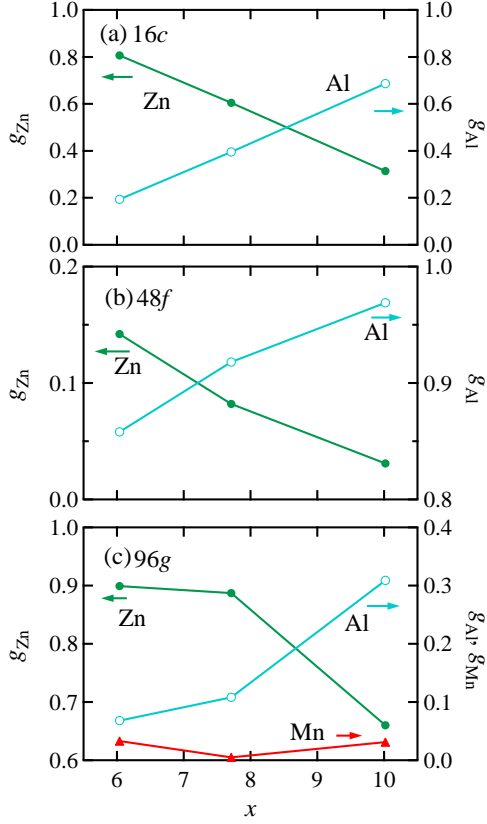}
\end{center}
\caption{(color online) Occupancies of the 16$c$, 48$f$, and 96$g$ positions in the Al-substituted samples. They are determined by $g$ listed in Table 7, assuming that the excess Mn atoms occupy the only 96$g$ site. The left axes represent the occupancy of Zn atoms, and the right axes represent those of Al and Mn atoms. 
}
\label{F13}
\end{figure}

The occupancies of Zn, Al, and Mn atoms in the 16$c$, 48$f$, and 96$g$ positions were estimated on the basis of chemical analyses and structural refinements, as shown in Fig. 13. The site preferences are also shown in Fig. 7. It is clear that Al atoms are present at all three sites and that their occupancies increase monotonically with increasing $x$. Note, however, that the Al occupancy in the 48$f$ position is relatively large and reaches 97\% for $x$ = 10.02, implying that Al atoms prefer this site. On the other hand, Zn atoms prefer the 96$g$ position: the occupancy of Zn atoms retains large values of $\sim$90\% for $x$ = 6.04 and 7.71, but decreases to $\sim$70\% for $x$ = 10.02. This change may affect the occupation of excess Mn atoms at the 96$g$ position. The difference in the site preferences of In and Al atoms may reflect the difference in their atomic radii in metals, i.e. In $>$ Al $>$ Zn; a large In atom prefers a large polyhedron around the 16$c$ position, while a small Al atom prefers a small polyhedron around the 48$f$ position. 

\begin{figure}[tb]
\begin{center}
\includegraphics[width=8cm]{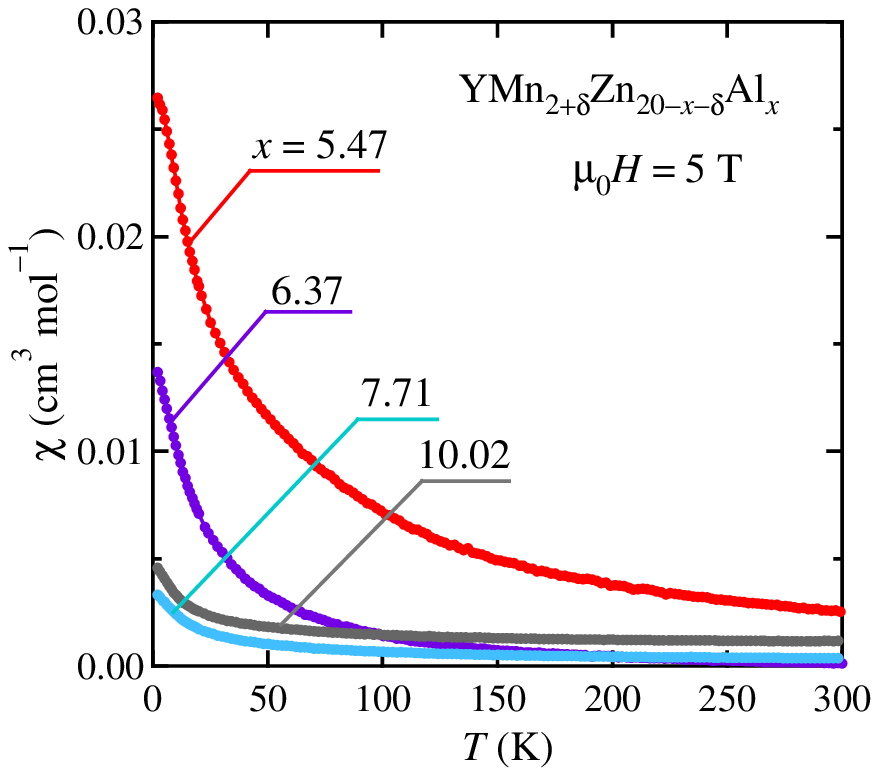}
\end{center}
\caption{(color online) Temperature dependence of magnetic susceptibility $\chi$ at a magnetic field of 5 T for Al-substituted samples.}
\label{F14}
\end{figure}

\subsubsection{Magnetic Susceptibility}
\label{S3-2-3}

Figure 14 shows the temperature dependences of magnetic susceptibility $\chi$ for four samples from $x$ = 5.47 to 10.02. The $x$ = 5.47 sample ($\delta$ = 0.59) shows the largest $\chi$ among the four samples, which is characterized by a Curie-Weiss curve with a Curie constant $C$ = 0.79 cm$^3$ K mol$^{-1}$ and a Weiss temperature $\theta_{\mathrm{W}}$ = $-$8.5 K. The value of $C$ is almost half that of the In-substituted samples with similar $\delta$, suggesting that a localized moment on an excess Mn atom is significantly reduced compared with that in the In-substituted samples. Moreover, antiferromagnetic interactions between the excess Mn spins are smaller, because the Weiss temperature is smaller than that in the In-substituted samples. 

As $x$ increases, magnetic susceptibility decreases and shows a less pronounced temperature dependence. Almost Pauli-paramagnetic behavior is observed for $x$ = 7.71 with $\delta$ = 0.06. The magnetic susceptibility of the $x$ = 10.02 sample shows similar Pauli-paramagnetic behavior, despite the fact that this sample contains a considerable amount of excess Mn of $\delta$ = 0.37. These results suggest that the magnetism of both excess and pyrochlore Mn atoms in the Al-substituted samples is considerably weaker than that in the In-substituted samples. We will discuss this issue later. 

\section{Discussion}
\label{S4}

As mentioned in the introduction, aluminum compounds $AB_2$Al$_{20}$ exist only for $B$ = Ti, V, and Cr to the left of Mn in the 3$d$ series, while zinc compounds $AB_2$Zn$_{20}$ exist only for $B$ = Fe, Co, and Ni to the right of Mn. Thus, $AB_2C_{20}$ compounds with $B$ = Mn can be prepared only when $C$ = Zn/In or Zn/Al~\cite{12}. We are interested in the physical properties of the embedded Mn atoms in the pyrochlore lattice. Unfortunately, however, it is found that the substitutions always introduce some amounts of excess Mn atoms that exhibit strong magnetism and tend to hinder probing of the magnetism of the pyrochlore Mn atoms. Hence, it is important to understand the crystal chemistry of the substitutions and to clarify the complexity in physical properties caused by the coexistence of the two kinds of Mn atoms. A possible future strategy may be to reduce the excess Mn and study in more detail the magnetism of the pyrochlore Mn atoms. 

\subsection{Crystal Chemistry of In- and Al-Substituted YMn$_2$Zn$_{20}$}
\label{S4-1}

First, we discuss the structural characteristics of In- and Al-substituted YMn$_2$Zn$_{20}$. The dependences on $x$ of the lattice constant $a$ are plotted in Fig. 15. The slopes are positive and larger for In than for Al, reflecting the atomic radii of In and Al, which are much larger and slightly larger than that of Zn, respectively, as mentioned above. Interestingly, the lattice constants obtained by extrapolating the two linear fits to $x$ = 0 almost coincide with each other, being estimated as 14.07(3) \r{A} and 14.07(2) \r{A} for In and Al substitutions, respectively. This confirms the reliability of our chemical analyses. Moreover, the values at $x$ = 0 must correspond to the lattice constant of pure YMn$_2$Zn$_{20}$, which has never been prepared. The estimated lattice constant of YMn$_2$Zn$_{20}$ is close to those of related compounds: $a$ = 14.10 \r{A} for YFe$_2$Zn$_{20}$ and $a$ = 14.054 \r{A} for YCo$_2$Zn$_{20}$~\cite{2,Jia}. 

\begin{figure}[tb]
\begin{center}
\includegraphics[width=7cm]{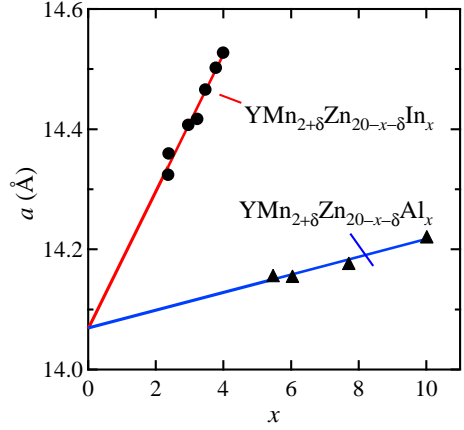}
\end{center}
\caption{(color online) Dependences on $x$ of the lattice constant $a$ for In- and Al-substituted samples. Two linear fits approach nearly the same value of 14.07 \r{A} at $x$ = 0, which should correspond to the lattice constant of pure YMn$_2$Zn$_{20}$. 
}
\label{F15}
\end{figure}

\begin{figure}[!htb]
\begin{center}
\includegraphics[width=6cm]{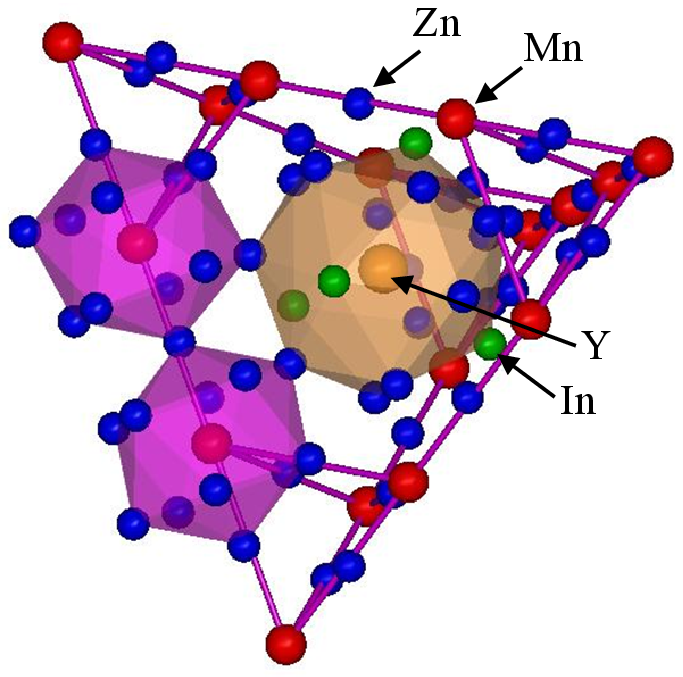}
\end{center}
\caption{(color online) Crystal structure of a hypothetical compound ``YMn$_2$Zn$_{18}$In$_2$'' with In and Mn atoms only in the 16$c$ and 16$d$ positions, respectively.}
\label{F16}
\end{figure}

The incorporated In and Al atoms are found to exhibit different site preferences over the three Zn sites of YMn$_2$Zn$_{20}$. The kinds of atoms occupying each site are shown in Fig. 7. In atoms preferentially occupy the 16$c$ position, which has the largest coordination number and distances to ligands, because an In atom has a much larger radius than a Zn atom. No more than two In atoms per formula unit can be accommodated at the 16$c$ position, and so any excess In occupies the 96$g$ position in the medium-sized polyhedron. At the same time, smaller Mn atoms tend to be incorporated at the 96$g$ position to compensate the size mismatch; an In atom is too large compared with a Zn atom to fit the polyhedron and thus may attract medium-sized Mn atoms. Thus, if an $x$ = 2 sample could be prepared, one could obtain the ideal compound ``YMn$_2$Zn$_{18}$In$_2$'' shown in Fig. 16, which is free from site disorder and magnetic contamination caused by the random distribution of In and Mn atoms at the 96$g$ position. 
\begin{figure}[tb]
\begin{center}
\includegraphics[width=7cm]{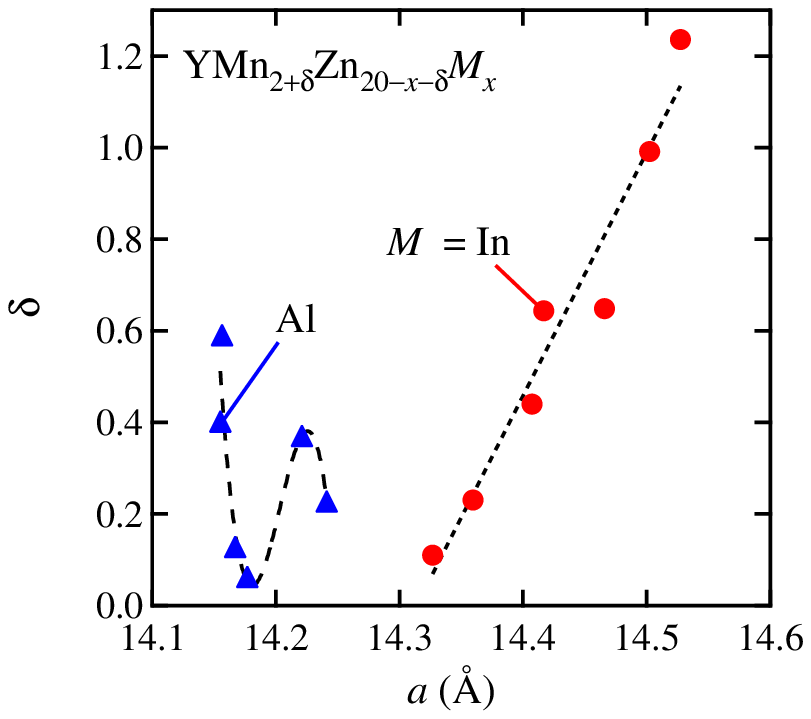}
\end{center}
\caption{(color online) Excess Mn content $\delta$ as a function of the lattice constant $a$ for In- and Al-substituted samples. The dotted line is a linear fit to the dataset of the In-substituted system. The broken line for the Al-substituted samples is a guide to the eyes.
}
\label{F17}
\end{figure}

\begin{figure}[tb]
\begin{center}
\includegraphics[width=6cm]{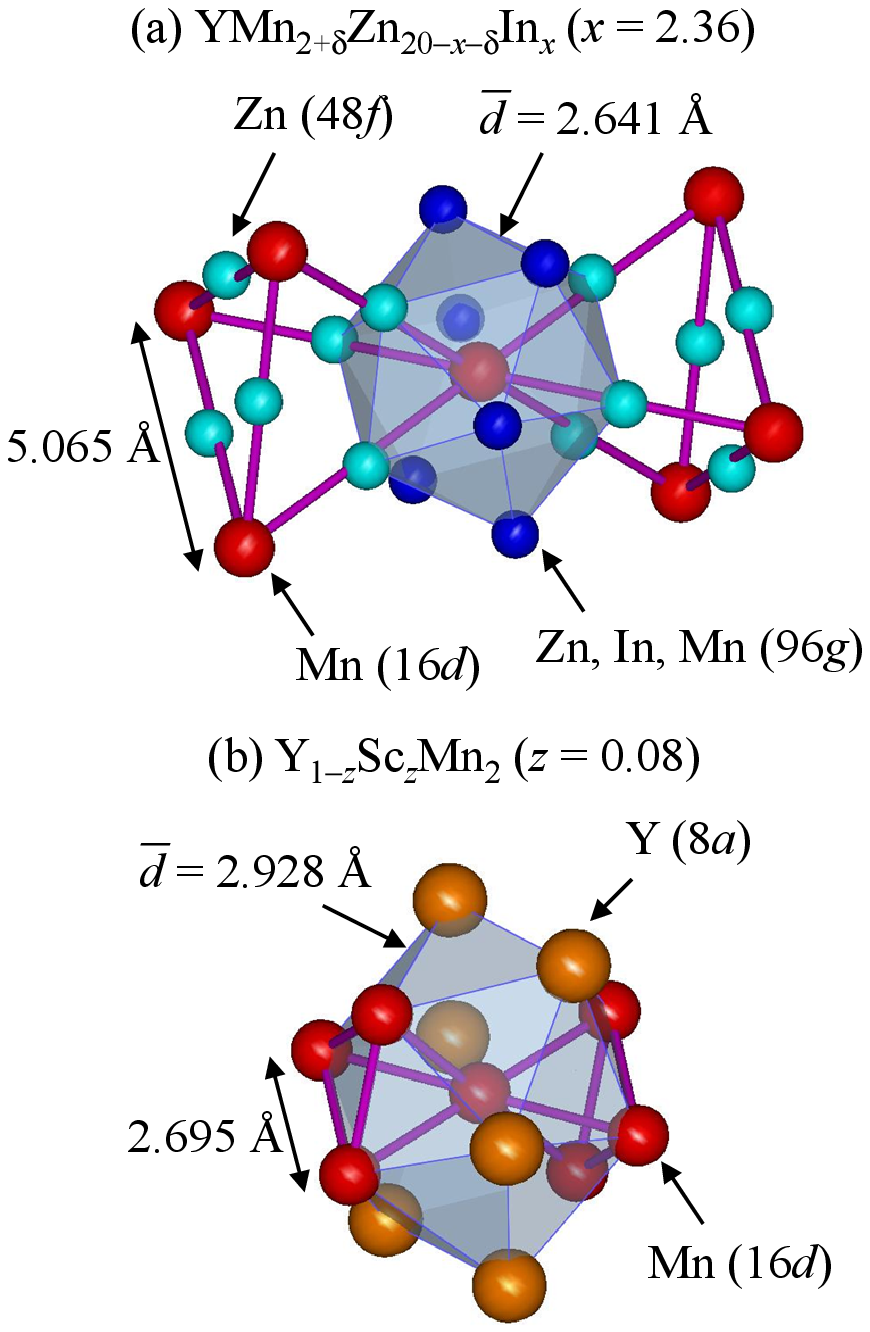}
\end{center}
\caption{(color online) Pyrochlore lattices comprising Mn atoms at the 16$d$ position and coordination polyhedra surrounding the 16$d$ position in In-substituted YMn$_2$Zn$_{20}$ (a) and (Y,Sc)Mn$_2$ (b)~\cite{8,YMn}. The pyrochlore lattices are represented by thick lines, and icosahedral coordination polyhedra are represented as semi-transparent polyhedra. The Mn-Mn distances and the averaged bond length from Mn to the ligands $\bar{d}$ are presented. 
}
\label{F18}
\end{figure}

In contrast to the In substitution, Al atoms partially occupy each of the 16$c$, 48$f$, and 96$g$ positions. This is probably due to the small difference in atomic radius between Al and Zn. However, as shown in Fig. 13, Al atoms tend to prefer the 48$f$ position. The different site preferences of incorporated In and Al atoms must have effects on the magnetism of Mn atoms, which will be addressed below. 

The presence of more magnetic, excess Mn atoms is undoubtedly a critical factor governing the physical properties of the present compounds. To study the properties of the pyrochlore Mn atoms, it is necessary to obtain a sample free from excess Mn. Let us consider here some possibilities. Figure 17 shows the excess Mn content $\delta$ as a function of lattice constant $a$ for the In- and Al-substituted compounds. In the case of In substitution, it seems that $\delta$ becomes 0 at $a$ $\sim$ 14.3 \r{A}, which corresponds to $x$ = 2 in Fig. 15. Thus, an $x$ = 2 sample would be free from both excess Mn atoms and site disorder caused by the random distribution of In atoms at the 96$g$ position, as discussed above. We have tried to prepare a sample with this composition, but have not yet been successful. Further optimization of preparation conditions would make it possible to obtain this ideal sample. 

In the Al-substituted system, $\delta$ tends towards 0 at around $a$ = 14.2 \r{A}, and then increases with increasing $a$. Hence, one strategy to remove the excess Mn in the Al-substituted system might be to prepare a sample with $a$ = 14.2 \r{A}. As shown by the linear fit in Fig. 15, this $a$ is realized at $x$ $\sim$ 9. Nevertheless, this sample would still be subject to disorder caused by the random distribution of Al atoms in the Zn sites.

\subsection{Magnetism of Mn atoms}
\label{S4-2}
\subsubsection{Difference between Pyrochlore and Excess Mn Atoms}
\label{S4-2-1}

The 3$d$ electrons of Mn atoms at the 16$d$ position forming a pyrochlore lattice are considered to be itinerant with weak antiferromagnetic correlations, i.e., $J$/$k_{\mathrm{B}}$ = 7 K for In substitution, while those from excess Mn atoms at the 96$g$ position behave as large local magnetic moments that also interact with surrounding spins through strong antiferromagnetic interactions, i.e., $J$/$k_{\mathrm{B}}$ = 49 K for the In substitution. This difference in magnetic character between the two types of Mn atoms can be understood by considering atomic distances between a Mn atom and its ligand atoms. Both the 16$d$ and 96$g$ positions are coordinated by twelve ligand atoms, as shown in Fig. 7. The bond length $d$ between a center and a ligand atom is smaller for the 16$d$ position than for the 96$g$ position. For example, in the In-substituted sample with $x$ = 2.96, averaged bond lengths to the twelve ligands are 2.641 and 2.875 \r{A} for the 16$d$ and 96$g$ positions, respectively; the difference between them is large, approximately 10\%. Since the smaller bond length means a larger hybridization, Mn 3$d$ electrons at the 16$d$ position have more itinerant character, while those at the 96$g$ position tend to be localized. 

In the In-substituted compounds, the excess Mn atom has a large magnetic moment comparable to that of spin 5/2, while in the Al-substituted compounds, it has a smaller moment for small $x$ and no magnetic moment for large $x$. This is probably because the average bond length for the 16$d$-ligands is smaller in the latter than in the former, which causes a larger hybridization for excess Mn atoms in the Al-substituted compounds. 

\subsubsection{Magnetic Properties of the Pyrochlore Mn Atoms and Heavy-Mass Electrons}
\label{S4-2-2}

Finally, we discuss the magnetism of the pyrochlore Mn in YMn$_2$Zn$_{20}$. As indicated by the Curie-Weiss fits to magnetic susceptibility in Fig. 9, pyrochlore Mn atoms in the In-substituted system each have an effective moment of about 1 $\mu_{\mathrm{B}}$ and interact with each other through weak antiferromagnetic interactions. However, the magnitude of this magnetic moment is significantly smaller than the large magnetic moment of 2.7 $\mu_{\mathrm{B}}$ observed in (Y,Sc)Mn$_2$~\cite{Nakamura}. This difference can also be rationalized in terms of the Mn-ligand distance. As compared in Fig. 18, the Mn-ligand distance in the In-substituted system is 10\% shorter than that in (Y,Sc)Mn$_2$. The shorter Mn-ligand distance can give rise to a stronger hybridization between Mn 3$d$ and ligand orbitals and thus reduce the magnitude of magnetic moments. In addition, the magnetic interaction in In-substituted YMn$_2$Zn$_{20}$ is small, $\sim$10 K, compared with several hundred K in (Y,Sc)Mn$_2$. This is because the size of the pyrochlore lattice is very different: In-substituted YMn$_2$Zn$_{20}$ has a Mn tetrahedron almost twice as large as that in (Y,Sc)Mn$_2$, as compared in Fig. 18. 

The Al-substituted samples show weaker magnetism than their In-substituted counterparts. In particular, those with large $x$ exhibit Pauli-paramagnetic-like magnetic susceptibility, indicative of the nonmagnetic nature of the pyrochlore Mn atoms, as observed in other $AB_2C_{20}$ compounds. It is likely that the itinerant character of the Mn 3$d$ electrons is enhanced by Al substitutions due to a larger hybridization between Mn and Al orbitals. Hence, In substitution may be more suitable for studying the magnetism of the pyrochlore Mn in YMn$_2$Zn$_{20}$ than Al substitution. 

The In-substituted YMn$_2$Zn$_{20}$ has antiferromagnetically interacting spins on the transition metal atoms forming a pyrochlore lattice, similar to those in (Y,Sc)Mn$_2$ and LiV$_2$O$_4$, which are $d$-electron heavy-fermion systems. Thus, one would expect a certain magnitude of mass enhancement in the In-substituted YMn$_2$Zn$_{20}$ derived from the suppression of magnetic order due to frustration. In the cleanest sample of $x$ = 2.36, in fact, $C_p$/$T$ increases rapidly toward $T$ = 0 and reaches 200 mJ K$^{-2}$ mol$^{-1}$, as shown in Fig. 10. This can be readily understood in terms of the enhancement of electronic heat capacity associated with the formation of a heavy fermion state. However, the presence of 5\% excess Mn atoms makes it difficult to carry out further quantitative analyses. Efforts to obtain a better sample are still in progress. 

\section{Conclusion}
\label{S5}

We have studied the In- and Al-substituted YMn$_2$Zn$_{20}$ systems by chemical analysis, structural refinement, magnetic susceptibility, and heat capacity measurements, and have shown that they have a general chemical formula of YMn$_{2+\delta}$Zn$_{20-x-\delta}$$M_x$ ($M$ = In, Al). Two kinds of Mn atoms with different magnetic properties are found: one kind forms a pyrochlore lattice, while the other excess Mn atoms are preferentially distributed over one of the Zn sites (96$g$). In the In-substituted case, the pyrochlore Mn atom has a small spin with an effective moment of approximately 1 $\mu_{\mathrm{B}}$, while the excess Mn atom has a large localized moment comparable to that of spin 5/2. In spite of antiferromagnetic interactions, neither of them shows any magnetic long-range order at low temperature. The absence of order in the pyrochlore Mn may give rise to a large enhancement in electronic heat capacity, indicative of the formation of a heavy-fermion-like state, similar to those observed in (Y,Sc)Mn$_2$ and LiV$_2$O$_4$, which are typical pyrochlore-lattice itinerant-electron antiferromagnets. In Al-substituted YMn$_2$Zn$_{20}$, magnetism originating from the Mn 3$d$ electrons is weaker than that in the In-substituted samples. 

To study in more detail the magnetic properties of the pyrochlore Mn atoms and the $d$-electron heavy-fermion state in YMn$_2$Zn$_{20}$, it will be necessary to prepare a cleaner sample with fewer excess Mn atoms. YMn$_2$Zn$_{18}$In$_2$ may possibly be an ideal compound with neither excess Mn atoms nor site disorder. 

\section*{Acknowledgments}
\label{S6}
We thank K. Ishida, R. Kadono, and H. Harima for helpful discussion. We also thank Y. Kiuchi for her help in carrying out the chemical analyses. This work was partly supported by a Grant-in-Aid for Scientific Research on Priority Areas ``Novel States of Matter Induced by Frustration'' (No. 19052003) provided by the Ministry of Education, Culture, Sports, Science and Technology, Japan. 



%% References


%\bibliographystyle{model1a-num-names}



\end{document}